\documentclass[manuscript]{aastex}
\usepackage{graphicx}
\usepackage{natbib}
\usepackage{txfonts}

\def\msun{$M_\odot$}
\def\mj{$M_{\rm J}$}
\def\etal{\emph{et al.}}

\begin{document}
\title{Interaction of a giant planet in an inclined orbit
       with a circum-stellar disk.}

\author{F. Marzari}
\affil{Universit\`a di Padova, Dipartimento di Fisica, via Marzolo 8, 35131
Padova, Italia}
\email{francesco.marzari@pd.infn.it}

\author{Andrew F. Nelson}
\affil{HPC-5, Los Alamos National Laboratory, Los Alamos NM, 87544, USA}
\email{andy.nelson@lanl.gov}

\begin{abstract}

We investigate the dynamical evolution of a Jovian--mass planet
injected into an orbit highly inclined with respect to its nesting
gaseous disk. Planet--planet scattering induced by convergent
planetary migration and mean motion resonances may push a planet into
such an out of plane configuration with inclinations as large as
$20^\circ-30^\circ$. In this scenario the tidal interaction of the
planet with the disk is more complex and, in addition to the usual
Lindblad and corotation resonances, it involves also inclination
resonances responsible of bending waves. 

We have performed three--dimensional hydrodynamic simulations of the
disk and of its interactions with the planet with a Smoothed Particle
Hydrodynamics (SPH) code. A main result is that the initial large
eccentricity and inclination of the planetary orbit are rapidly damped
on a timescale of the order of $10^3$ yrs, almost independently of the
initial semimajor axis and eccentricity of the planet. The disk is
warped in response to the planet perturbations and it precesses.
Inward migration occurs also when the planet is inclined and it has a
drift rate which is intermediate between type I and type II migration.
The planet is not able to open a gap until its inclination becomes
lower than $\sim 10^\circ$ when it also begins to accrete a
significant amount of mass from the disk. 

\end{abstract}

\keywords{stars: planetary systems --
        planetary systems: formation
                       }

\section{Introduction}

Planet migration has been recognized as a key mechanism of evolution
for planetary systems. Torques on a planet, arising from interactions
with the circumstellar disk, can cause a planet to move a considerable
distance from the orbit where it formed into a new dynamical
configuration, possibly explaining the many observed giant planets
moving on trajectories well within the ice condensation line and very
close to the parent star. Unfortunately, it cannot explain the highly
elliptical orbits of many of the extrasolar planets discovered to
date. The most likely explanation for these objects is that another
mechanism comes into play in the early phases of evolution of a
planetary system: planet-planet scattering. When two or more planets
form from the protoplanetary disk, they may come very near each other
and begin a period of chaotic evolution dominated by mutual close
encounters. In such a scenario, one (or more) of the planets can be
ejected from the system entirely, leaving the remaining planets with
eccentric orbits. This mechanism, in its classical formulation
\citep{wmjj,marwe,rafo,chatfo}, is based on the assumption that the
planets emerge from the protoplanetary disk, after the dissipation of
the gaseous component, on orbits which are packed together closely
enough to be unstable. The subsequent evolution in a gas free scenario
dramatically alters the initial orbital structure of the multi--planet
system, ejecting bodies out of the system and causing the growth of
the eccentricity and mutual inclination of the surviving planets in
the system. 

Orbital migration by tidal interaction with the disk and by
planet--planet scattering are usually studied as separate mechanisms
occurring at different evolutionary stages of a planetary system.
However, it is likely that they acted together if the protostellar
disk does not dissipate shortly after the formation of Jovian planets.
Recent numerical models by \citet{alimo} show that giant planet
formation may require only $10^5-10^6$ yr to occur, although many
uncertainties remain in understanding the process. As a consequence, a
multi--planet system may already be formed while the bodies are still
embedded within the disk. The subsequent type I or type II migration
\citep{ward} may lead any pair of planets to come closer than their
stability limit and start a planet--planet scattering phase. The
chaotic evolution dominates the subsequent evolution of the planets
and the outcome is not significantly different from the gas--free
scenario. At the end of the scattering phase, if the gas component of
the disk is still present, the tidal interaction between the disk and
the surviving planets may still modify the final configuration of the
system by moving the planets close to the star or by circularizing
their orbits.

Planet migration after a planet--planet scattering phase has to
account for the high eccentric and inclined orbits the planets have
acquired during the chaotic evolution. So far, both analytic and
numerical work has studied migration of giant planets mainly in the
context of initially circular orbits coplanar to the protoplanetary
disk. \citet{moorad} considered two--planets systems evolving under
the action of disk torques (type II migration) and planet--planet
scattering. If the planets formed sufficiently close together or moved
close to each other during their formation process while they evolved
under the action of type I migration, which depends on the mass of the
body, their combined tidal effects would cause of the region between
them to be cleared because of overlapping Lindblad resonances.
Hydrodynamic simulations \citep{snell,Papa,kley1,kley2,NB03a,NB03b}
show that the timescale for clearing is only a few hundred orbital
periods and also that in many cases the region inside the inner planet
orbit quickly loses material due to accretion onto the host star. The
inner planet does not migrate because there is no disk material left
close to it to generate torques. On the other hand, the outer disk
exerts negative torques at the Lindblad resonances of the outer planet
possibly causing it to migrate inward moving closer to the inner one.

Convergent orbital migration due to different drift rates may cause
planets to end up either in resonance or to move into crossing
trajectories that eventually start a chaotic scattering phase, in turn
frequently causing a planet to be ejected. The surviving planet is
left on an eccentric and inclined orbit that is evolved, in the model
of \cite{moorad}, with the usual migration rate of a planar orbit.
However, if the planet orbit is out of the disk plane, its inward
orbital migration may be significantly different from the planar case.
The interaction between the planet's gravitational field and the gas
is complicated by the dynamical effects related to the inclination
between the disk and the planet. The planet might be unable to clean
and maintain a gap in the disk, but still generate waves at Lindblad
and vertical resonances and cause torques that dominate the dynamical
evolution of the planet. The high eccentricity that a planet typically
acquires after the chaotic phase might also reverse the migration
trend away from the star, as shown by \cite{papl}. Disk warping in
response to the secular perturbations of the planet might lead,
through dissipation, to damping of the inclination \citep{lubo},
though we note that his conclusions are derived under the assumption
that the planet opens a gap in the disk, splitting it in two rings.
These are all aspects that need a detailed investigation from a
numerical point of view.

An additional mechanism that can excite the inclination of a giant
planet embedded in a disk is capture in mean motion resonance.
\citet{tholi} have shown that two Jupiter--mass planets on converging
orbits may enter inclination resonances during their evolution. These
resonances induce rapid growth of the inclinations of both planets
which end up in highly non--coplanar configurations. According to
their simulations, once the planets are trapped in a 2:1 mean motion
resonance, their eccentricities are excited while they migrate
together towards the star. During this evolution the planets may enter
a 4:2 second order inclination--type resonance whose critical
arguments include the node longitudes. The inclinations of both bodies
increase rapidly and the inner planet may achieve an inclination
larger than $\sim 20^{\circ}$ in some tens of thousands of years,
while the eccentricity of both bodies keep growing. After a while, the
system moves out of resonance with a further abrupt increase in
inclinations. The important condition for this resonance to occur is
that the mass ratio between the inner planet and the outer planet is
not smaller than two. Otherwise the inner more massive body cannot
gain enough eccentricity to enter the inclination--type resonance.
Another critical aspect of the resonant evolution is the amount of
eccentricity damping by the disk. If it is too strong, it might
prevent the planets from reaching the critical eccentricity that opens
the door for inclination--type resonances.  

In this paper we concentrate on the orbital evolution of a giant
planet injected into a highly inclined, highly eccentric orbit respect
to the disk. This may have occurred either after a planet--planet
scattering phase or once captured in an inclination--type resonance
and escaped from it. In both cases we expect not only a high
inclination but also a large eccentricity. We will investigate the
kind of orbital migration the planet undergoes and the timescale of
its eccentricity and inclination damping. Previous studies
\citep{tawa,pala} focused on the 3--dimensional interaction between
small planets and the disk with eccentricities and inclinations
smaller than the disk aspect ratio. In this regime the effects of
density and bending waves on the planet evolution have been
analytically computed with a linear model, with \citet{tawa}
predicting damping times for both eccentricity and inclination of the
order of $300 (r/1 AU)^2)$. \citet{cress} numerically simulated the
evolution of small mass planets (up to 20 Earth--mass) on orbits with
eccentricities up to $e=0.3$ and a maximum inclination relative to the
disk of $10^{\circ}$ via 3--dimensional hydrodynamical simulations,
beyond the range of validity of the \citet{tawa} analytical model.
From their fit to the data in a high eccentricity ($e=0.3$) and high
inclination regime ($i=8^{\circ}$), the damping time of both $e$ and
$i$ to low values is approximately of the order of 400 orbital periods
($\sim 4700$ yr) for a planet at 5.2 AU. They also always find inward
migration, even if reduced for large $e$ and $i$. 

Here, we consider 3D simulations of massive planets (one Jupiter mass)
evolving on highly inclined ($20^{\circ}$) and eccentric ($e=0.4$)
orbits. The scenarios are different since the evolutionary history
leading to the high inclination orbit is distinct. Moreover, the
migration mechanism is expected to be significantly different due the
large mass we consider compared to that analysed in previous models.
The linear regime adopted by \cite{tawa} to model density and bending
waves does not hold in the case we consider and, in addition, the disk
is also significantly warped in response to the planet perturbations. 

In Section 2 we describe the initial conditions and numerical
techniques while in section 3 we describe the results of the
simulations. In Section 4 we comment our results and discuss future
work.

\section{Initial conditions and physical model}

The systems we simulate consist of three components: a star, a planet
and a disk. We model these systems using the publically available
code, VINE, \citep{vineI,vineII} to model the combined system in three
spatial dimensions, using the particle based `Smoothed Particle
Hydrodynamics' (SPH) method. This code has previously been used by one
of us in extensive simulations of circumstellar disks
\citep[e.g.][]{NBR00,Nelson06} and is known to perform well in such
configurations. 

We define the star and planet to have mass 1\msun\ and 1\mj,
respectively, and to be in Keplerian orbits around their common center
of mass. The orbit is eccentric, with $e$ ranging from 
$0.0$ to $0.4$ in different simulations, and inclined respect
to the disk plane (defined below) by $i = 20^{\circ}$. We consider
three different initial semi-major axes for the planet: 2, 4 and 6 AU.
These orbital parameters can easily be achieved either after a phase
of planet--planet scattering or because of resonant pumping as
described in \citet{tholi}. The star and disk are each modeled as
Plummer softened point masses, with softening lengths of 0.25~AU and
0.05~AU, respectively. Self gravity for the disk, when included in our
simulations, is calculated using an approximate, tree based summation
of particles and aggregates of particles, with parameters set such
that the maximum force errors are smaller than $\sim0.1$\%. 

We define the initial condition of the disk as follows. At time zero,
we set approximately 450000 equal mass particles on a series of
concentric rings extending from an innermost ring at a distance of
0.5~AU from th star, to an outer radius at 13~AU. Disk matter is set
up on initially circular orbits assuming a Keplerian rotation curve.
Radial velocities are set to zero. The distribution of disk material
is defined by a surface density power law of the form
\begin{equation}\label{eq:denslaw}
\Sigma(r) = S \Sigma_0 \left[ 1 + \left({r\over r_c}\right)^2\right]^
{-{p\over{2}}},
\end{equation}
where $\Sigma_0=15500$~gm/cm$^2$, $p=3/2$ and $r_c=0.25$~AU. These
parameters yield as disk mass of approximately 0.0078\msun, or just
over 8\mj. In order to avoid numerical instabilities in the initial
condition associated with a too sharp outer disk edge, we include a
softening function, $S$, in the specification of the surface density,
as originally specified in \citet{Nelson06}. This function extends
$\pm\delta=1$~AU in each direction from the nominal outer disk edge,
and alters the surface density using a function which decreases
linearly to zero over the range, so that the disk's final dimensions
extend to 14~AU. The distribution of particles in the vertical
coordinate is specified below.

The temperature profile is specified by a similar power law, of the
form:
\begin{equation}\label{eq:templaw}
T(r) = T_0\left[1 + \left({r\over r_c}\right)^2\right]^{-{q\over{2}}},
\end{equation}
where $T_0=340$~K, $q=1/2$ and $r_c$ is as above. With these
parameters, the `snow line', at which planet formation is expected to
be enhanced, falls at $r\approx3.9$~AU. We employ a locally isothermal
equation of state to close the system of hydrodynamic equations, with
temperature at any given radius always specified exactly by equation
\ref{eq:templaw} and the pressure specified by the relation $p=\rho
c_s^2$, where $c_s$ is the sound speed. This choice permits a
computationally simple description of the disk thermodynamics at the
cost of restricting the physical interpretation given to the
simulations. In some circumstances for example, it may permit gas to
compress to much greater densities than would otherwise be the case.
We refer the reader to \citet{DBMNQR_PP5} for a more complete
discussion. These restrictions will incur no great burden on our
calculations except in the neighborhood of the planet, where they may
become inaccurate, particularly when appreciable disk material
collects there as it does at late times in our simulations. We must
therefore be cautious in our interpretations of the behavior at such
times.

The choice of an isothermal equation of state constrains the vertical
density structure of the disk to be a Gaussian function of the $z$
coordinate, with a scale height, $H=c_s/\Omega$ \citep[see
e.g.][]{Nelson06}. The specification of the initial condition will be
complete once we incorporate this constraint. We do so by assigning a
value to the $z$ coordinate for each particle using the locally
defined isothermal scale height, multiplied by pseudo-random number
drawn from a Gaussian distribution. 

With the initial conditions described above, the value of the well
known Toomre stability parameter is well above $Q=20$ everywhere in
the disk. Therefore, we expect the system to be stable to self
gravitating disturbances. Nevertheless, self gravity may remain an
important effect because of its influence on the radial positions of
resonances in the disk \citep{NB03b}, which in turn affect the
strength of gravitational torques on the planet and its migration.  

As a final check on our initial condition, we have verified that our
simulations obey minimum resolution requirements, below which the
numerical methods fail to reproduce known behavior of the system
accurately. Following the procedure described in \citet{Nelson06}, we
have generated a replica of our initial condition without a planet,
and evolved it for 300~yr in order to permit a stable numerical
equilibrium condition to evolve for the vertical structure. We then
performed empirical fits to the vertical disk structure as realized in
this simulation, and compared the parameters obtained from these fits
to their analytically expected values. This exercise demonstrates that
our simulations are adequately resolved, with ratios of fit to
analytic midplane densities above 90\% over the radius ranges of
interest to our calculations. 

\section{Results}

Using the initial conditions and physical model described above, we
have run a family of similar simulations of the evolution of a planet
initially in an orbit inclined to a circumstellar disk. In the
following sections, we describe the evolution of a prototype model,
then a number of variations upon it. Our prototype model is of a
planet initially orbiting its parent star with a semi-major axis of
4~AU, eccentricity $e=0.4$, and with its orbital plane inclined by
$i=20^{\circ}$ to a circumstellar disk, for which we neglect the self
gravitating effects of the disk on itself. One variation studies
changes in evolution when disk self gravity is included. Two consider
the identical initial condition, realized at higher and lower
numerical resolution. Two study the changes in evolution when the
planet's initial semi-major axis is either 2 or 6~AU. Two study
changes in the evolution when the planet's initial eccentricity is
reduced to $e=0$ and $e=0.2$. Finally, we study evolution of the
planet's motion when the disk has an internal cavity, into which a
planet is assumed to have been scattered on to an orbit with
semi-major axis of 2~AU.

\subsection{Evolution of systems with and without disk self gravity}

Self-gravity may play an important role in the evolution of a planet
interacting with the disk. A number of previous works (\cite{NB03a,
NB03b, bama}) have shown that the inclusion of disk self--gravity in
numerical hydrodynamical models lead to a different value of the
differential Lindblad torque and may indeed affect the computed
migration rate of a planet embedded in the disk. For this reason we
performed two simulations with the same initial parameters for both
the planet and the disk, but in one case we took into account disk
self-gravity. 

\subsubsection{Orbital evolution of the planet}

\begin{figure}[ht!]
\epsscale{0.1}
\includegraphics[angle=-90,scale=0.55]{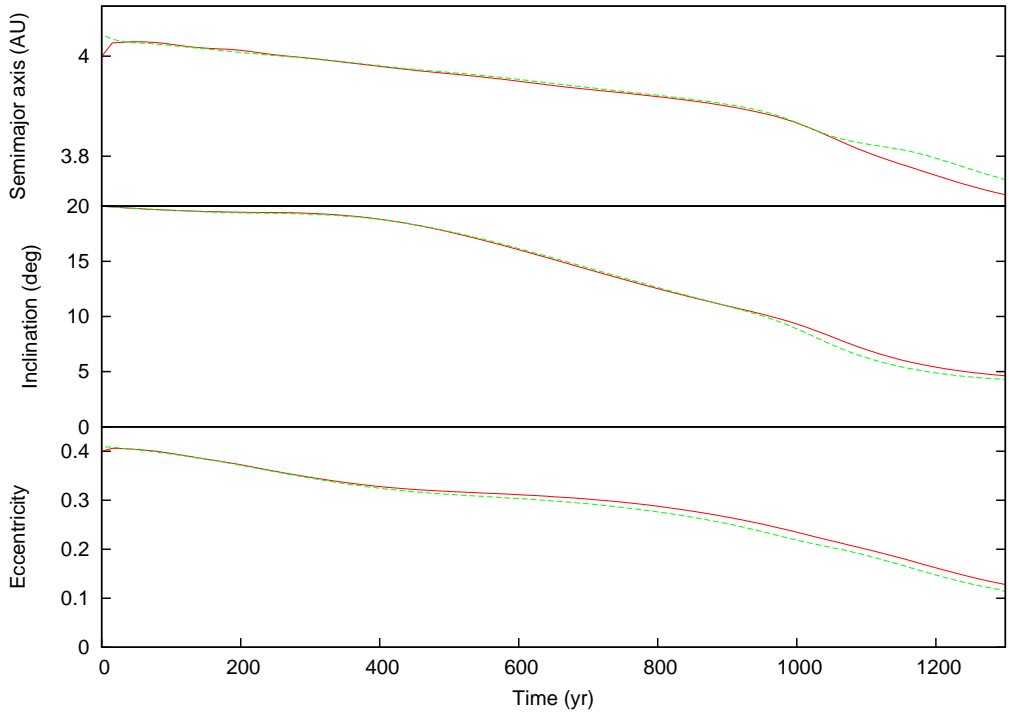}
\caption{\label{fig:sgr} 
Evolution of the orbital elements of planets residing in disks with
(continuous line) and without (dashed line) self gravity. From top to
bottom, we show semi-major axis, eccentricity and inclination as
functions of time.} 
\end{figure}

In figure \ref{fig:sgr} we show the evolution of the planet orbital
elements with the planet started on an orbit with semimajor axis
$a=4$AU, eccentricity 0.4 and inclination $20^\circ$. 
The exchange of angular momentum between the disk and planet causes
all of the planet's orbital elements to evolve significantly over the
$\sim1000$~yr time scale we simulate here, in both simulations.
Overall, both simulations show essentially identical behavior, and we
conclude that disk self gravity is not an important effect in systems
like those in our study. Therefore, in the interests of numerical
simplicity, will neglect it in our remaining discussions.

As the systems evolve, the planet undergoes inward orbital migration
at a rate of $\sim1.5$~AU per 10$^4$~yr, consistent with analytic
theories of Type~I migration \citep{tana}, but for planet of
terrestrial mass. Due to the large initial inclination of the planet,
it is expected that not only Lindblad and corotation resonances cause
the disk-planet coupling but that also vertical resonances
\citep{demur,griv,tawa} play a significant role. These resonances have
the node longitude of the planet and that of the ring at the resonant
location within their resonant argument. It is well known that the
existence of vertical resonances with Mimas give rise to bending waves
in the Saturn rings. The combination of eccentric and bending waves,
in particular in the initial phase of the planet evolution when it is
well out--of--plane, makes it difficult to predict analytically all
the torques acting on the planet. In addition, and as we will show
below, the disk becomes warped and precesses, adding further
complexity to the problem. Our scenario is significantly different
from that described by \cite{tawa} since we consider a Jovian type
planet and values of eccentricity and inclination by far larger than
the disk aspect ratio. Therefore, we limit ourselves to give the
numerical value of the migration rate. 

Most significantly in the context of our interest in the outcomes of
planet--planet scattering, both the planet's eccentricity and its
inclination are quickly damped by its interaction with the disk.
Although the interaction strength between disk and planet are clearly
highly time dependent at early times as the planet successively dives
through the disk midplane and climbs out of it during its orbit, both
inclination and eccentricity decay at near constant rates until
relatively late in the simulations. The decay accelerates
significantly after $\sim800-1000$~yr of evolution, as the planet
spends more and more time close to the midplane where densities are
high. At approximately the same time, they begin to accumulate an
envelope of disk material because of their longer duration passages
through the midplane. We will discuss the numerical and physical
significance of the envelope mass accumulation in section
\ref{sec:atmo}. 

\subsubsection{Evolution of the disk morphology}

\begin{figure}[htp]
\epsscale{0.1}
\hskip -3 truecm
\includegraphics[angle=-90,scale=0.8]{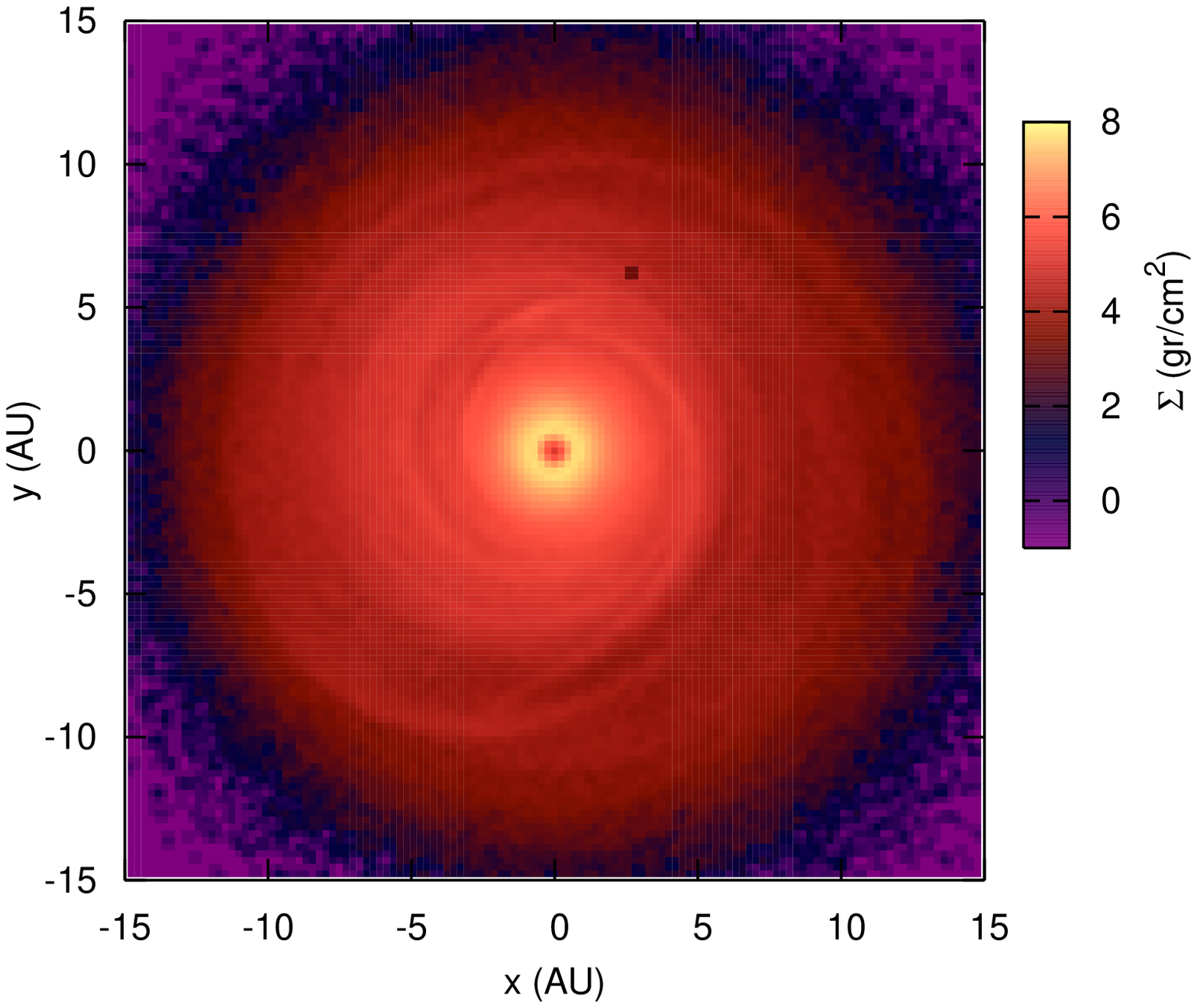}
\hskip -0.8 truecm
\includegraphics[angle=-90,scale=0.8]{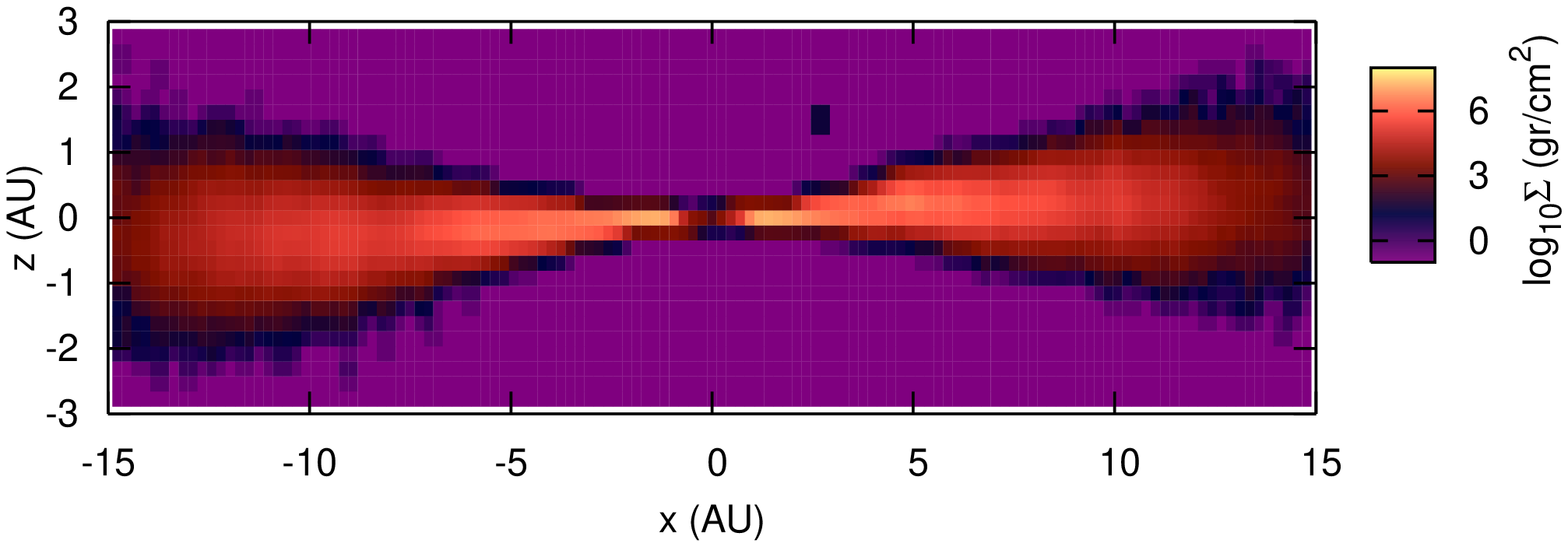}
\caption{\label{fig:disk}
Density distribution of the disk after 500 yrs from the beginning of
the simulation where the disk is unperturbed. In the top plot we show
the logarithm of the disk density ($gm/cm^2$) on the (x,y) plane while
an angular section of the disk projected in the (z,y) plane is shown
in the bottom plot. The position of the planet is marked by a black
square in both plots.} 
\end{figure}

Figure \ref{fig:disk}a shows the disk surface density morphology for the
simulation with self gravity, projected onto the $(x,y)$ plane after
500~yr. Spiral structures generated by the tidal interactions between
planet and disk extend inward and outward from it's position,
demonstrating the effect of Lindblad resonances on the evolution. No
low density gap has yet developed at this time, nor does one develop
later, until the planet's trajectory decays to an orbit where it again
spends most of its time embedded in high density disk material.

\begin{figure}[htp]
\epsscale{0.1}
\includegraphics[angle=-90,scale=0.90]{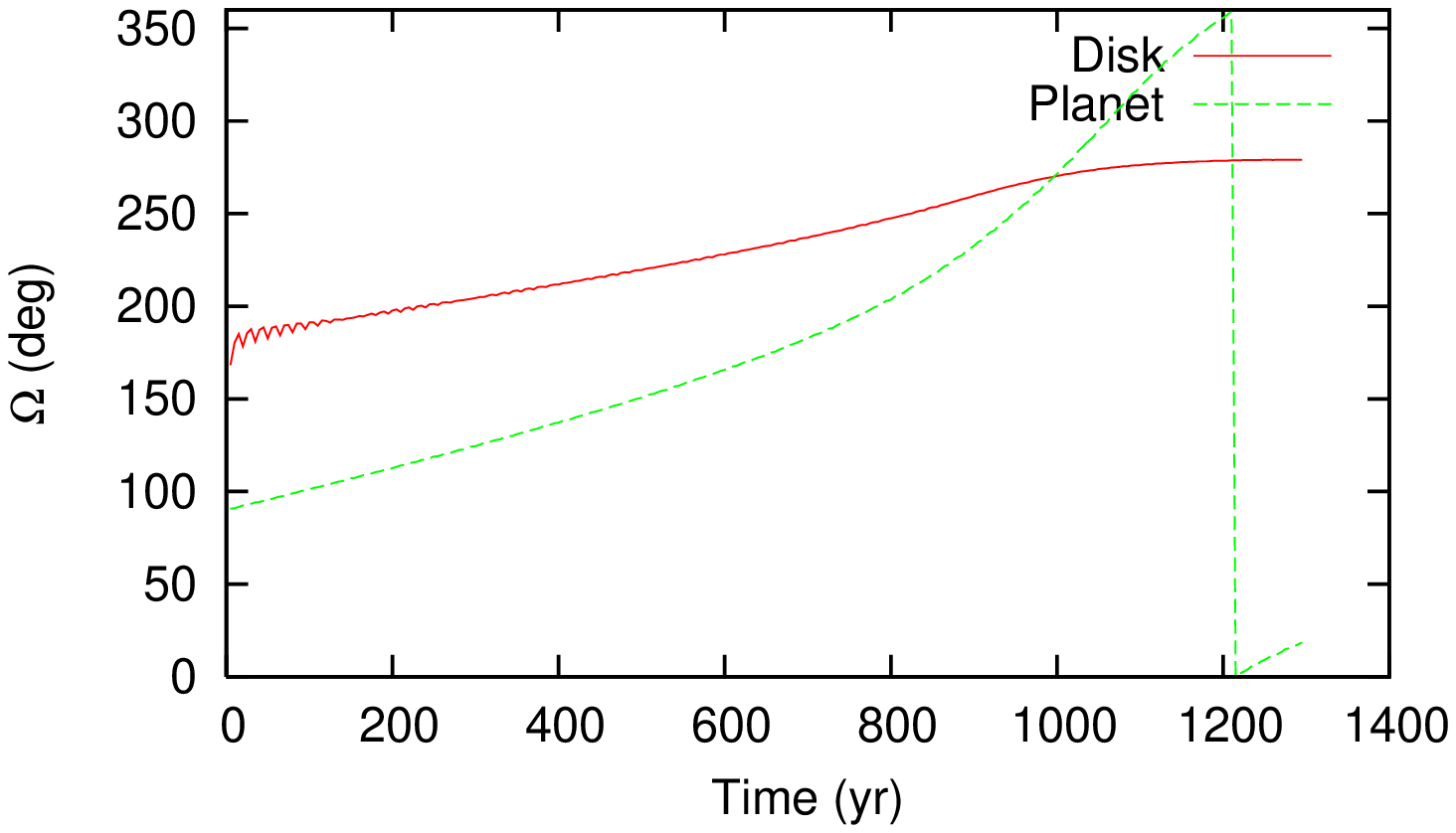}
\caption{\label{fig:node}
Evolution of the node longitude of the disk (continuous line) and
planet (dashed line). The precession of the disk is slower as the
planet dives into the disk on low inclination orbits.} 
\end{figure}

The out-of-plane perturbations of the planet leads to a warped
structure of the disk, as shown in the projection plot of Figure
\ref{fig:disk}b, and to its precession. The shape of the disk in the
(x,z) plane is similar to that found by \cite{larw}, who modeled the
evolution of a disk in a binary system not coplanar to the companion
orbit. Our scenario is similar to theirs, with the planet being less
massive than a binary companion but closer to the star and the disk.
In figure \ref{fig:node} we show the evolution of the node longitude
of both the planet and the disk, respect to the starting reference
frame, when the planet has an initial semimajor axis of 6 AU. The
precession of the disk is fast when the planet is far from the disk
plane and it slowly decreases when the orbit of the planet becomes
less inclined. 

\subsection{Evolution of systems realized with different numerical
resolution}

\begin{figure}[ht!]
\epsscale{0.1}
\includegraphics[angle=-90,scale=0.55]{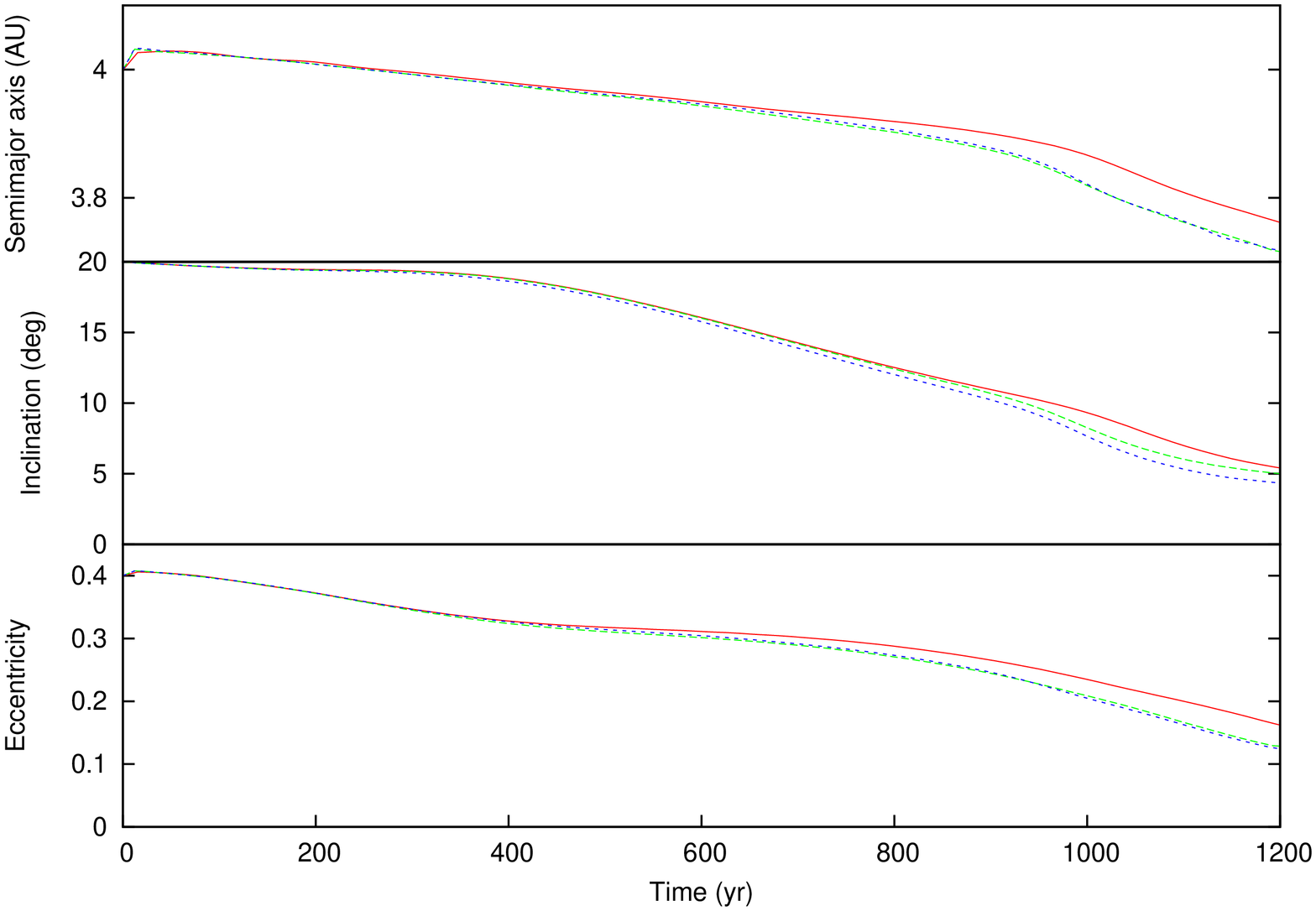}
\caption{\label{fig:reso-vary} 
Evolution of the orbital elements of planets residing in disks
resolved at low (blue dashed line), standard (red continuous line) and high
(green dashed line)  numerical resolution, as defined in
the text.
Each panel shows the same quantity as in figure \ref{fig:sgr}.} 
\end{figure}

In order to investigate the sensitivity of our simulation results to
numerial resolution, we compare the the orbital evolution of otherwise
identical simulations, differing only in numerical resolution from our
non-self gravitating prototype model. One simulation uses half the
number of particles used in our prototype ($\sim225000$) and a second
uses twice that number ($\sim900000$). Figure \ref{fig:reso-vary} shows
the evolution of the planet's orbital elements for all three
simulations. The evolution of the orbital elements is nearly identical
in all three simulations, until $\sim900$ have passed. Given this
outcome, we believe that the evolution of the orbital elements seen in
our simulations is well resolved.

At late times, the planet's inward migration accelerates in all three
simulations, but much more so in the low and high resolution variants
than at standard resolution. We believe the difference is due to a
slightly more permissive condition placed upon the smoothing length in
the low and high resolution simulations, than was placed on the
standard resolution variation. Specifically, a larger minimum value of
the smoothing length for particles was employed for the standard
resolution run than the other two. As the planet reenters the disk and
began accreting mass from it, smaller smoothing lengths permit
comparatively more massive envelopes to develop, resulting in larger 
perturbations to the dynamical evolution in the low and high
resolution simulations. We will discuss the significance of the
envelope in our simulations in section \ref{sec:atmo}, below.

\subsection{Evolution of systems with varying initial orbital
parameters}

\begin{figure}
\epsscale{0.1}
\includegraphics[angle=-90,scale=0.55]{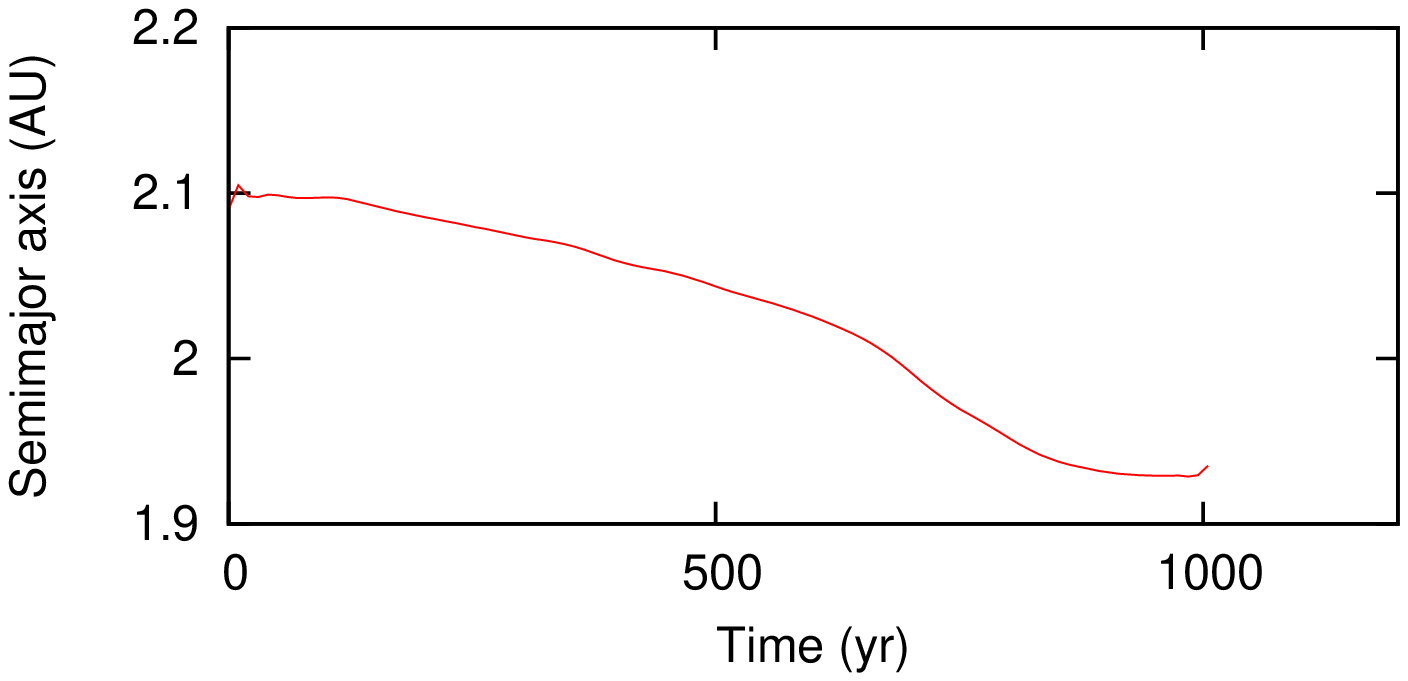}
\includegraphics[angle=-90,scale=0.55]{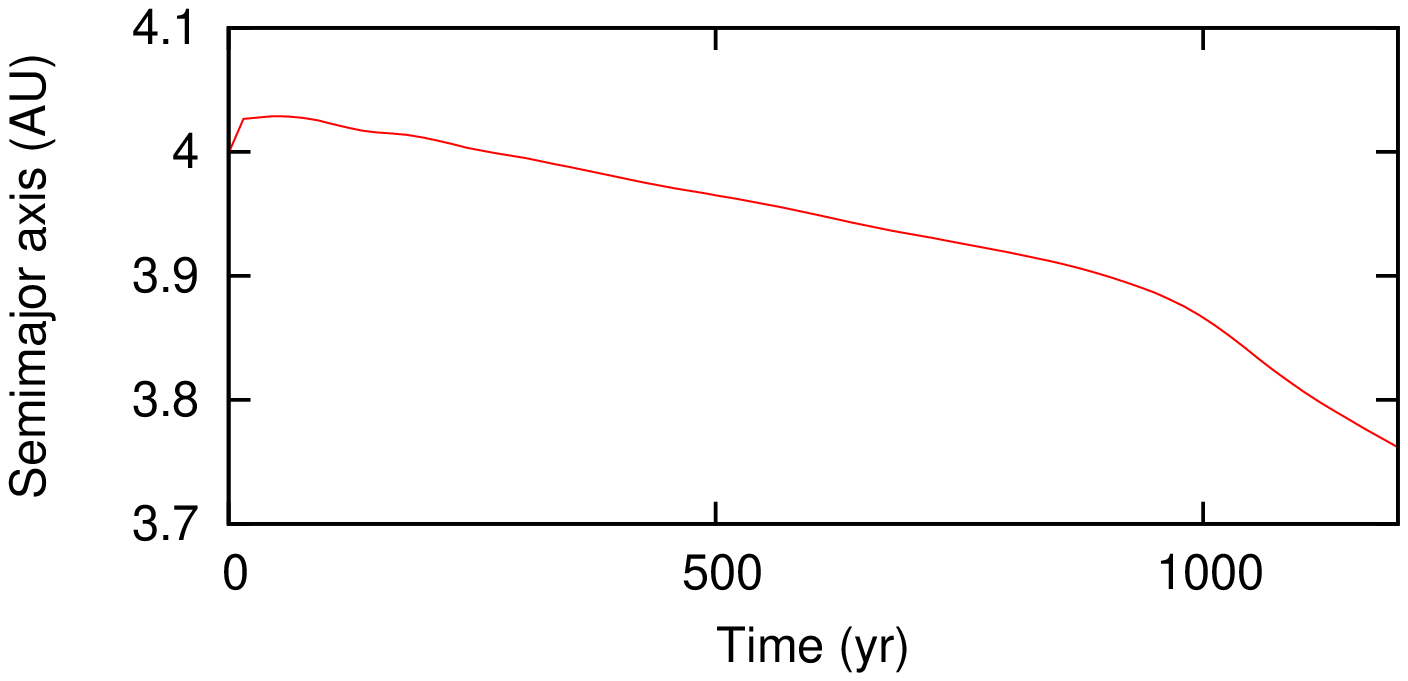}
\includegraphics[angle=-90,scale=0.55]{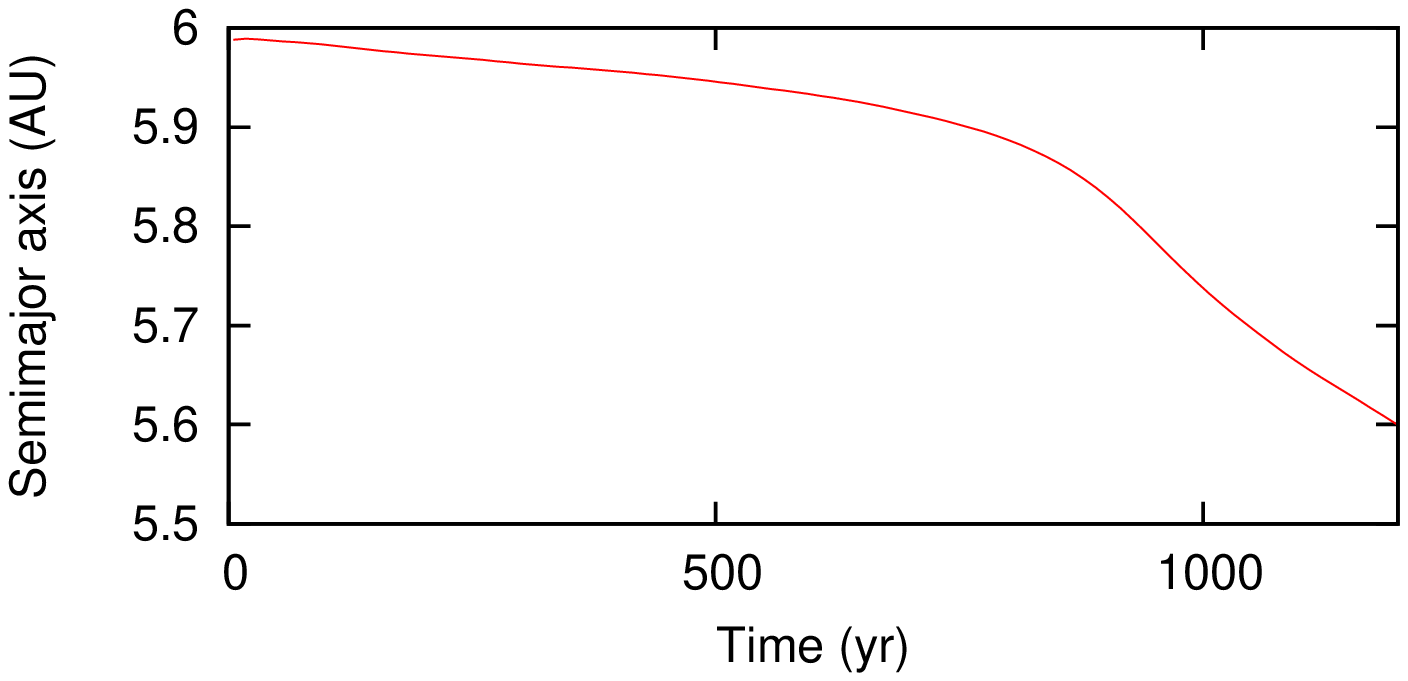}
\caption{\label{fig:elements1}
Semimajor axis evolution of the planet as a function of time for three
different initial semi-major axes, near 2, 4 and 6 AU. Apparent sharp,
short term variations in the semi-major axes are artifacts due to our
calculation of its instantaneous value based on the planet's position
and velocity parameters at the time of each dump.}
\end{figure}

Figure \ref{fig:elements1} shows the migration of the planet for three
different initial semimajor axes. In each case, the initial migration
phase is characterized by similar inward drift rates which do not
significantly depend on the initial location of the planet and also do
not significantly depend on the osculating eccentricity and
inclination. After the orbits are damped to sufficiently low
inclination--about 10$^\circ$ in each case--migration rates increase
because the planets spend more time embedded in the disk where density
is high. As noted above for the two 4~AU models, the planets also
begin to accumulate an envelope of disk material at this time. 

\begin{figure}
\epsscale{0.1}
\includegraphics[angle=-90,scale=0.55]{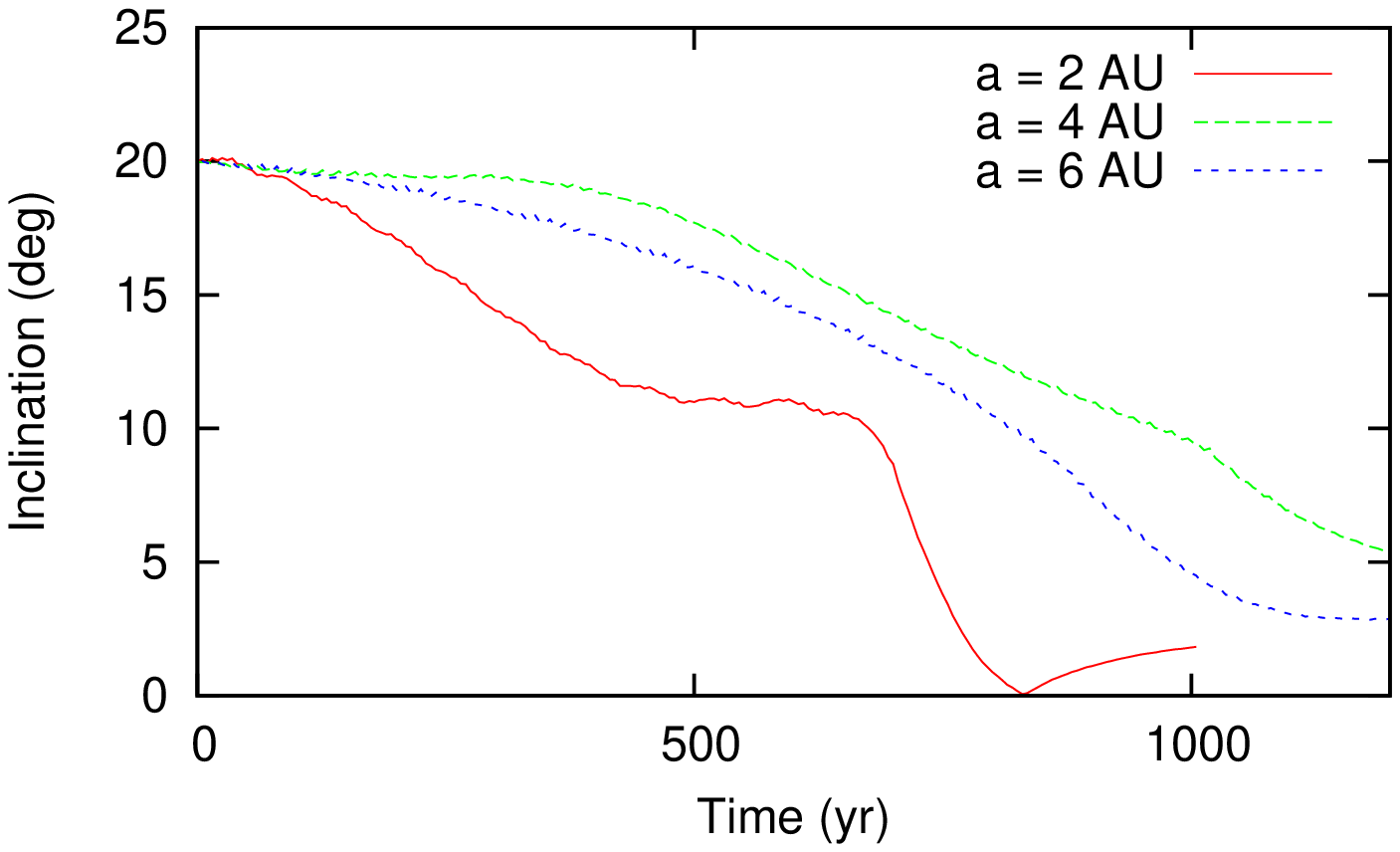}
\includegraphics[angle=-90,scale=0.55]{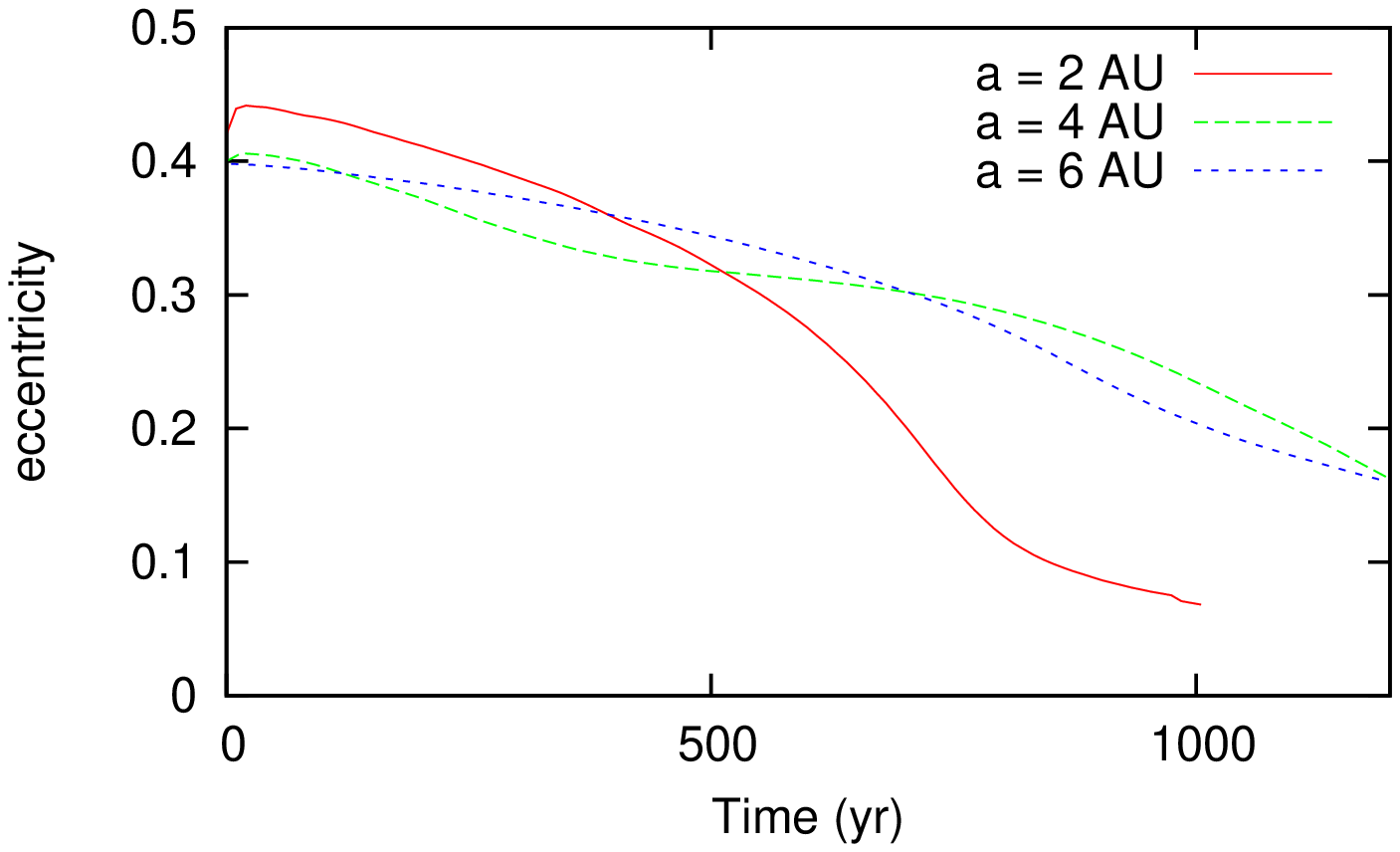}
\caption{\label{fig:elements2}
Evolution of a planet's eccentricity and inclination. Both the orbital
elements are damped by the interactions of the planet with the disk. }
\end{figure}

Both eccentricity and inclination are damped rapidly as the disk and
planet interact, with rates that depend slightly on the initial
semimajor axis of the planet (Fig. \ref{fig:elements2}. The damping is
similar to that observed by \citet{cress} for high eccentric orbits.
In the first phase the decay is described approximately by $ \dot{e}
\propto e^{-2}$ while it becomes exponential when $e$ is smaller than
0.1, as predicted by the linear analysis of \citet{tawa}. This last
kind of evolution is really observed only for the case with $a_0 \sim
2$ AU, since the inclination damping is faster in the other two cases
($a_0 \sim 4, 6$ AU) and the planet re-embeds itself entirely inside
the disk before the eccentricity becomes lower than 0.1. Performing a
least--squares fit to the data with a function like that given in
\citet{cress}, we find that the damping time down to $e = 0.1$ is 265
orbital periods, $T_0$ for $a_0 \sim 2$ AU, 154 $T_o$ for $a_0 \sim 4$
AU and 103 $T_o$ for $a_0 \sim 6$ AU. This is faster than that found
by \citet{cress} for a protoplanet with a mass of 20 $M_{\oplus}$
which is about 400 $T_0$ for $e_0 =0.3$, $i_0 = 8^\circ$, and for $a_0
\sim 5.2$ AU. A more massive planet evolves more rapidly because its
interaction with the disk is stronger. 

The inclination damping rates are more sensitive to the initial
location in the disk, with the 2~AU model decaying most rapidly.
Interestingly, after it begins to accumulate an envelope, the damping
increases dramatically. Ultimately the planet passes entirely through
the disk, such that its apoapse lies below the disk plane rather than
above it, visible in the plot near $\sim$~700~yr as an inflection
point in the value of the inclination. As it reaches low inclination
however, the eccentricity decay slows because essentially all disk
matter within reach of the planet's orbit has become fallen into its
envelope. Little mass remains to provide continuing damping through
continuing gravitational torques. Neither of the other two simulations
were evolved far enough in time to determine whether they too would
produce such behavior. The damping time of the inclination appears to
be slightly shorter than that of the eccentricity. 
As for the eccentricity, a
scaling proportional to $i^2$ can be used to model the initial
non--linear damping of inclination.  We find that the timescale for
damping down to $6^\circ$ ($\sim 0.1$ radian) with a least squared fit
is 212 $T_0$ at 2 AU (the fit is poor in this case), 122 $T_0$ at 4 AU
and 64 $T_0$ at 6 AU. 

\begin{figure}
\includegraphics[angle=-90,scale=0.55]{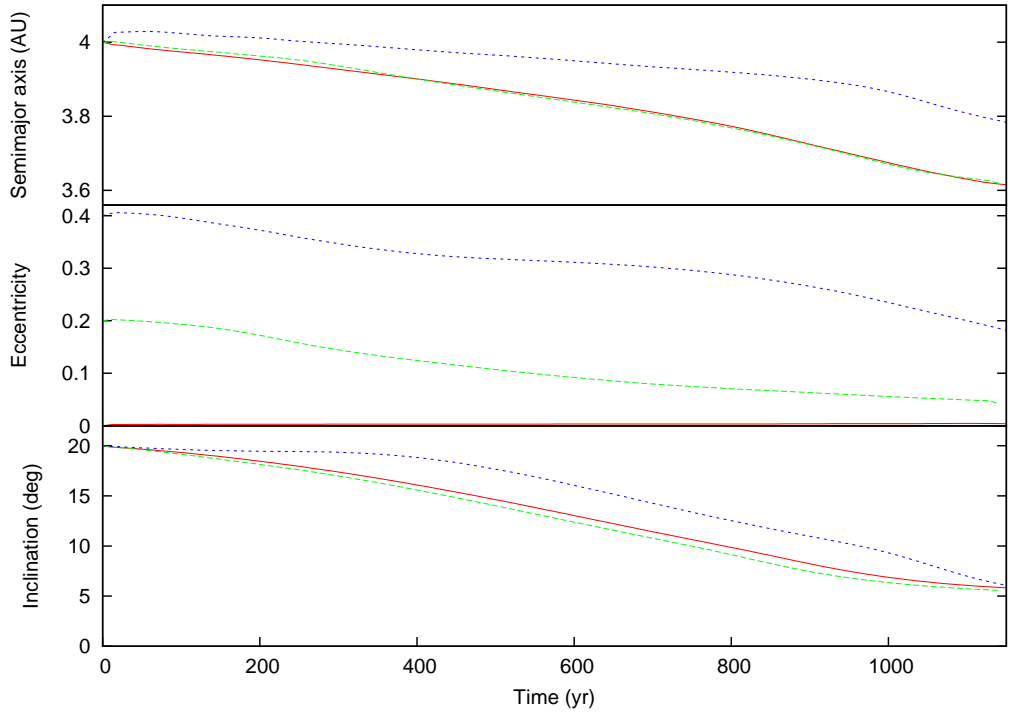}
\caption{\label{fig:ecce_dep}
Orbital evolution of a planet for three different initial values of
eccentricity. Both the migration rate and the inclination damping are
faster for lower eccentricities. The red line is for $e_0=0.0$, the
green line for $e_0=0.2$ and the blue line for $e_0=0.4$.}
\end{figure}

If we start the planet with a lower initial eccentricity both the
migration rate and the inclination damping are faster. As it can be
seen in \ref{fig:ecce_dep} the planet started with an eccentricity of
0.4 has a slower evolution compared to the cases with $e_0=0.2$ and
$e_0=0.0$. Only after the planet has developed an envelope and has
returned to the disk the orbital behaviour is almost independent from
the eccentricity. The slower evolution with higher eccentricity is
possibly related to the larger vertical distance from the disk at
aphelion. A planet on a high inclined and eccentric orbit spends more
time far from the disk compared to one on a circular orbit leading to
a lower planet--disk interaction. The effect is not linear with
eccentricity since the orbital evolution for both $e_0=0.2$ and
$e_0=0.0$ is quite similar while it is significantly different when
the eccentricity grows to $e_0=0.4$.

Based on the results of these simulations, we conclude that planets
excited onto moderately inclined and eccentric orbits will return to a
state in which they are embedded within the disk within a few thousand
years of the event which originally caused their orbit excitation.

\subsection{Evolution of systems where the planet has been scattered
into a disk cavity} \label{sec:cavi}

As discussed in the Introduction, a phase of planet--planet scattering
may be a consequence of convergent migration of planets. Hydrodynamic
simulations of two close planets embedded in the disk
\citep{snell,Papa,kley1,kley2,NB03a,NB03b} have shown that they may be
able not only to clear of disk material the region in between them but
also that inside the orbit of the inner planet forming a cavity. A
similar result is also obtained when 3 planets are considered
\citep{matsu}. When the planets are driven close enough by the outer
disk viscous evolution they may undergo either resonance trapping or
mutual scattering depending on the physical conditions of the disk. Ad
example, \cite{adams} have shown that ongoing convergence of orbits
leads to scattering when the disk is turbulent. A possible outcome of
the planet--planet scattering phase in this scenario is that one
planet is injected on an inner eccentric and inclined orbit within the
cavity while the other(s) are thrown on outer orbits. A similar
evolution for a system of 2 planets has been studied by \cite{moe}. 

Here we concentrate on a scenario in which a planet is injected into a
cavity of the disk, possibly created prior to the scattering phase,
where it evolves under the action of the residual outer disk.  We
start the planet on an orbit with a semimajor axis of 2 AU, an
inclination of $20^\circ$ and eccentricities of $e=0.0$ and $e=0.4$.
The disk parameters and the resolution of the numerical model are the
same as in the previous simulations but in this case we truncate the
disk inside a given radius. We consider two different models, one
where the disk is truncated at 4 AU and one at 5 AU. 

\begin{figure}
\includegraphics[angle=-90,scale=0.5]{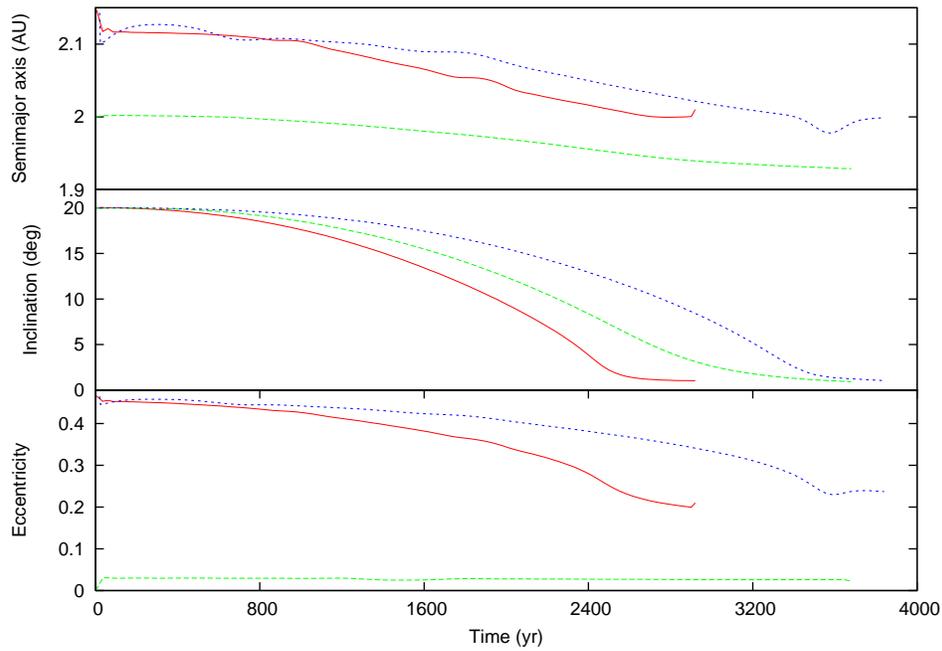}
\caption{\label{fig:trun-disk}
Orbital evolution of a planet scattered into a disk cavity on a high
inclination orbit ($i_p = 20^{\circ}$). The red solid line illustrates
the case where the initial eccentricity of the planet is 0.4 and the
disk is truncated inside 4 AU. The blue dashed line corresponds to a
truncation radius of 5 AU. The green dashed line shows the evolution
of a planet on an initially circular orbit embedded in a disk
truncated at 4 AU.} 
\end{figure}

In figure \ref{fig:trun-disk} we show the evolution of the orbital
elements of the planet. The inclination damping occurs similarly to
the case without truncation (fig. \ref{fig:elements2}) but the
timescale is longer. The planet returns to the disk plane after about
2500 yrs if the truncation radius is at 4 AU, and 3500 yrs with the
truncation at 5 AU. The inclination damping rate depends on the mass
of the disk which is affected by its gravity and the rate is reduced
when the disk is less massive. When the truncation is set to 4 AU the
mass of the disk is about 4.6\mj  and it is reduced to 3.98\mj  when the
truncation radius is 5 AU.  Even the orbital circularization occurs on
a longer timescale and the planet has still an eccentricity of about
0.2 when it falls back onto the disk. 

\begin{figure}
\includegraphics[angle=-90,scale=0.5]{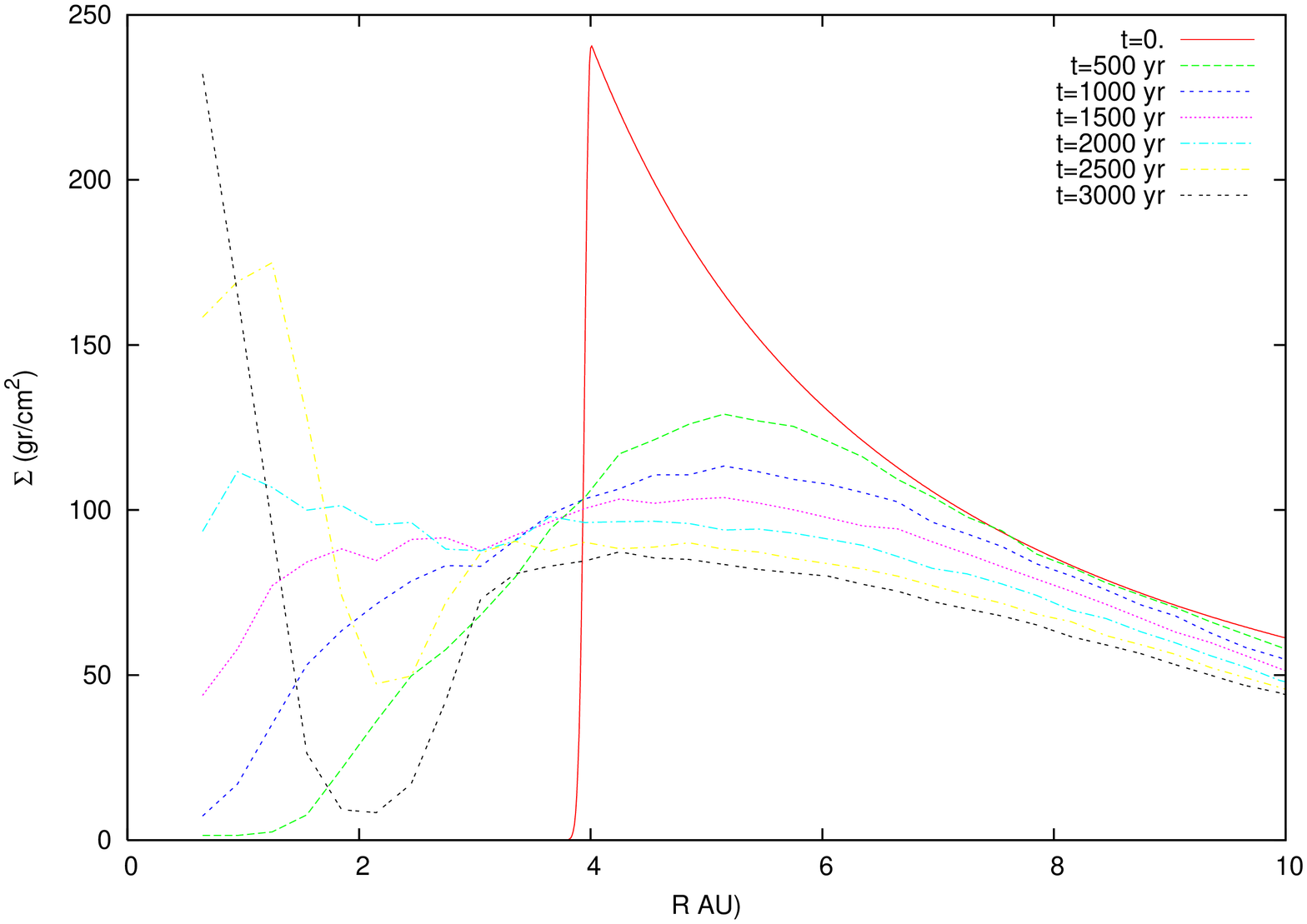}
\caption{\label{fig:profile} 
Density profile of the disk as a function of the radial distance
from the star at different evolutionary times. The gas refills the
inner region of the disk until the planet dives back into the disk 
and opens a gap. }
\end{figure}

The rates and overall magnitudes of semimajor axis migration are
consistently smaller compared to the case without truncation (fig.
\ref{fig:elements1}). The rates increase with time however, and we
postulate that they are related to the increasingly strong
interactions with disk material that slowly refills the cavity as the
evolution proceeds. The evolution of the disk density profile,
averaged over the azimuthal angle, is illustrated in Fig.
\ref{fig:profile} at different evolutionary times for the case with an
initial truncation radius of 4 AU. As time passes, gas flows past the
truncation radius and resupplies the inner region of the disk, due to
both hydrodynamic forces with the disk itself and gravitational
perturbations from the still inclined planet.

The cavity is replenished with gas coming from the outer disk only
while the planet remains on an inclined orbit. As soon as the planet
reaches the disk, it reopens a gap and it evolves as predicted by type
II migration. Similar behaviours are observed in both our simulations
with a broader initial cavity and those with the planet on an
initially circular orbit. The presence of a cavity or of a smaller
gap, possibly originating in the planet/disk interactions prior to the
planet--planet scattering phase, does not affect significantly the way
in which the inclination is damped, but may lead to an increase of the
timescale of the damping process. Even if some gas survives in the
cavity, as suggested by \cite{cri,luda}, this would not alter the kind
of planet evolution we have described. Again, only the timescale and
possibly the eccentricity with which the planet falls back into the
disk may be different.

\subsection{Effects of accretion and the planet's
atmosphere}\label{sec:atmo}

As a consequence of its presence in the disk, mass accretes onto a
forming planet and an envelope grows around it. During times when it
spends most of its orbit at high altitude above or below the disk
midplane, less mass will accumulate, since that part of its orbit is
spent in regions where densities are low. To what extent will
accretion continue, while still inclined? Can it be enhanced by
repeated encounters with the high density disk midplane or does
efficient accretion require the core to be permanently embedded?

From a theoretical standpoint, it is unclear whether a planet in a
highly inclined orbit should have a substantial envelope, accumulated
before its orbit was perturbed out of the midplane, or how massive
that envelope should be if present. If the high orbital inclination is
a consequence of a planet--planet scattering phase, the strong
gravitational interactions during the close encounters might have
stripped the original atmosphere away. On the other hand, scenarios
where high inclination is pumped up by a resonance would appear to be
much more complex. In this case, the planet would be gently pushed out
of plane over time and might retain its atmosphere.

Our models do not consider accretion of gas (e.g. via absorbing SPH
particles directly onto the planet), nor do they include either
sufficient resolution or a sufficiently detailed physical model for us
to accurately model the formation of an envelope around the planet
itself. Therefore, at times when accretion is significant, our results
will become less representative of the actual physical behavior than
when accretion is not important. With care however, we shall be able
to draw a number of interesting conclusions from our results during
both time frames.

\begin{figure}
\includegraphics[angle=-90,scale=0.5]{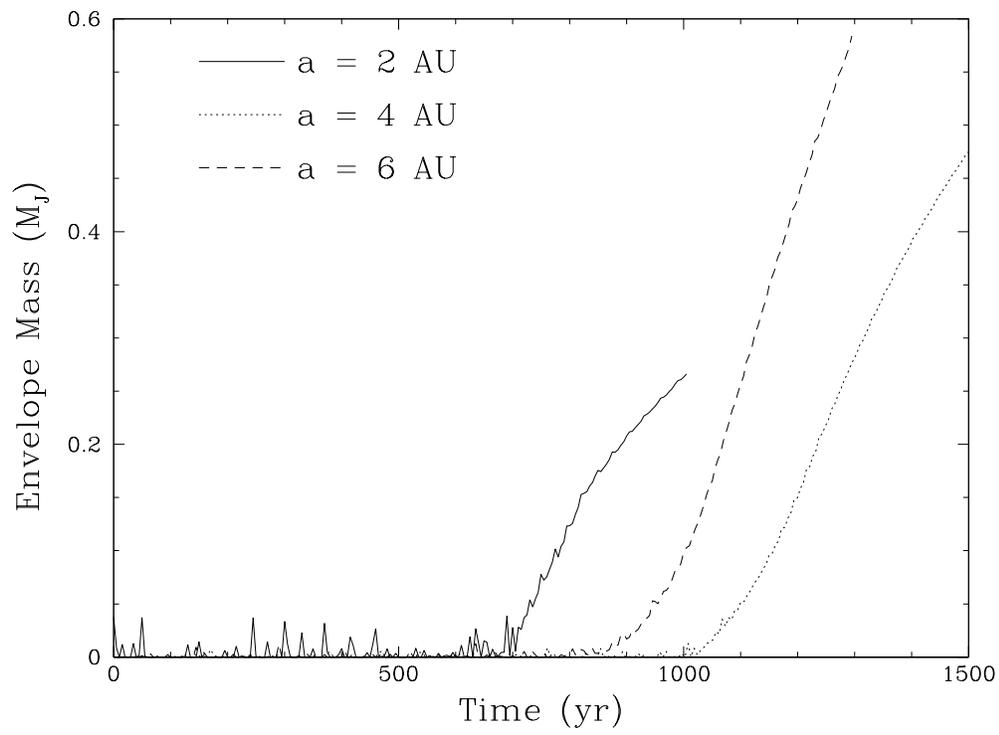}
\caption{\label{fig:env-mass}
Mass accumulation within a volume of one Hill radius in diameter for
our simulations with varying initial semi-major axis. }
\end{figure}

Figure \ref{fig:env-mass} shows the mass of material contained in an
envelope of one Hill radius, $R_H$, in diameter, for each of our three
simulations without disk self gravity. Near the beginning of each
simulation, disk material appears inside the Hill volume only when the
planet passes through the disk itself, and does not accumulate. Mass
begins to accumulate only after the orbit inclinations have decreased
to $\sim 10^\circ$, approximately 700~yr, 1000~yr and 900~yr after the
simulations themselves begin for the 2~AU, 4~AU and 6~AU runs,
respectively. Once it does, the envelope masses increase monotonically
to several tenths of the planet's mass within a few hundred years of
additional evolution.

In principle, the presence of an extended envelope is appropriate. It
corresponds to the situation in which the envelope is near hydrostatic
equilibrium and the gravitational contraction timescale (the
Kelvin--Helmholtz time) is longer than the evolutionary timescale
under consideration, in this case inclination and eccentricity
damping. This will occur when the envelope continues to accrete while
the central core of the planet continues to radiate strongly
\citep{polla,nelso}. As a consequence, any accumulated atmosphere does
not fall onto the planet because it is sustained by its own thermal
energy. 

In fact however, once accumulation begins, the rates derived from our
plots are extremely high, and so are instead more likely due to the
breakdown of the assumption that the gas is well modeled by a locally
isothermal equation of state. In the neighborhood of the planet, this
assumption fails because the gas undergoes substantial compression and
because it interacts dynamically with streams of material that enter
the Hill volume on different trajectories. Both processes generate
thermal energy very quickly, thereby increasing the fluid pressure and
throttling further accretion. Additionally, a planet on a
significantly inclined orbit will spend most of its time out of the
disk, thereby permitting it to be heated further by stellar
irradiation. Using an isothermal gas is equivalent to assuming that
all of this energy is instantly radiated away so that the pressure
remains low and throttling is short circuited, but violating the
constraints of Kelvin--Helmholtz contraction noted above.

The fact that essentially no material accumulates in the envelope
while its orbit remains inclined, even though our assumptions strongly
favor it, supports the physical conclusion that no further accretion
occurs during that time. For planets whose inclinations never decay
below a threshold of $\sim 10^\circ$, this statement is equivalent to
the conclusion that they have reached their final masses, perhaps long
before the disk itself has decayed away and even while other planets
in the system continue to grow. During this time period we believe
that our simulations most accurately reflect the actual evolution of
the orbital elements, because accretion plays no role.

Once accumulation begins, the large envelope masses perturb the
migration rates, such that after the envelope has formed they are
likely overestimates. There are at least two reasons of numerical
origin to support this conclusion. First, the slight differences in
the orbital velocities of the disk material composing the envelope and
the planet cause the envelope to exert a large effective drag force,
slowing the planet in its orbit and increasing its migration. A
similar phenomenon appears in the grid based simulations of
\citet{NB03a} when the gravitational softening parameter was set
smaller than approximately half of one zone, and we believe that the
origin of both behaviors is the same. 

Second, a numerical phenomenon similar in effect to friction may play
an important role in the evolution. As the envelope grows in size, so
too does its cross section for interacting with disk material. This is
important because the artificial viscosity used in SPH, required for
numerical stability, produces far more dissipation in disk simulations
due to shear than is appropriate disk simulations, a fact well known
to its practitioners. With a large cross section, the planet/envelope
combination suffers much greater dissipation than is necessary for
flow stability, thereby increasing the migration rates as orbital
kinetic energy dissipates in the form of heat. While we employ several
techniques to reduce the magnitude of the dissipation \citep[see
e.g.][]{vineI}, it is not possible to eliminate it entirely, so the
migration will therefore be more rapid than would otherwise occur. 

Due to these effects, we expect a slower migration rate and a slower
damping of eccentricity and inclination than actually occurs late in
our simulations, though still more rapid than the initial evolution
when the planet resides primarily outside the disk. Although we cannot
rely on the rates for changes in the orbital elements to be fully
accurate at late times, we are still able to draw important
conclusions from them because they are upper limits. First, we can
conclude that the migration of a planet on an inclined orbit will
accelerate once its inclination decreases to a value below $\sim
10^\circ$, where it again becomes embedded in the disk. Its
eccentricity and inclination will also be damped more quickly than
occurs at higher inclinations. Systems in which planet--planet
scattering has occured will therefore be observed in configurations
where the planets either remain in very different inclinations than
the original disk, or in essentially coplanar configurations because
the inclinations have been near totally damped.  

\section{Discussion}

When a Jovian planet is ejected out of its protoplanetary disk after a
phase of planet--planet scattering or because of resonant
interactions, it is rapidly pulled back into the disk. The
gravitational interaction between the disk and the planet causes a
quick dumping of both the eccentricity and inclination of the planet
orbit while, at the same time, the disk becomes warped. The timescale
by which the planet returns to the disk midplane is of the order of a
few thousand years, significantly faster than that previously found
for small planets ($M \leq 20 M_{\oplus}$). The interaction with the
gaseous disk also causes the planet to migrate inwards at a rate
faster than the viscous one and comparable to that of a terrestrial
planet under Type~I migration. Many different resonances may
contribute to this out of plane inward orbital migration, including
vertical resonances. The inclination damping is a robust result and
also occurs when the planet is injected into a disk cavity, where it
interacts with the outer residual disk after a presumed scattering
event. The damping rates depend on the amount of gas outside the
cavity and on details of how the gas refills the cavity after the
planet has been moved to its highly inclined orbit. Our results imply
that systems with multiple planets each with different inclination,
should have undergone planet--planet scattering after the disk was
already dissipated or that a different mechanism acted to stir the
inclinations. We can make a similar conclusion for systems in which
planetary orbits are misaligned with the stellar spin axis, assuming
that the present day stellar spin axis defines the disk plane during
formation.

During its evolution far from the disk midplane the planet accretes
mass at a very low rate. However, when its inclination is lower than
$\sim10^\circ$ its accretion rate increases rapidly as mass becomes
trapped within its Hill's sphere. This leads to a faster damping of
eccentricity and inclination and also to a more rapid inward
migration. Our SPH simulations do not model this last phase accurately
because the isothermal approximation fails when gas is accreted and
compressed in proximity of the planet and it also interacts with
streams of material entering the Hill's sphere on different
trajectories. 

The problem of the interaction of the disk with the growing atmosphere
of the planet is a very complex, beyond the aim of this paper. One
simple method to improve our numerical modelling in the last stages of
the planet evolution when it starts to accrete disk mass could be to
adopt an 'ad--hoc' accretion rate of the material within the Hill's
sphere, as implemented in numerous previous works. Unfortunately, the
details of the accretion prescription do not provide much physical
insight, as they must be chosen rather arbitrarily.

The scenario we outline with our model generates many questions in
need of further investigation. For example, during planet--planet
scattering in presence of the gas in the protostellar disk, how much
of the planet envelope is lost during mutual close encounters? Is the
tidal force strong enough to strip each planet of a significant amount
of gas? What is the thermodynamical state of the envelope after a
sequence of close encounters? 

As the planet returns to the disk, how quickly does the envelope
reform? Does it reform quickly into a hydrostatic configuration? How
massive? Does it trigger a revisit of the core instability phase
during which the planet originally gained most of its mass? Does
infalling disk material instead produce a highly dynamically active
envelope, as was seen for lower mass cores in \citet{NR05}?  

An additional interesting aspect is to understand what happens when an
additional giant planet is added to the system, either on an inclined
or planar orbit. Is the interaction between the disk and the inclined
planet different? Do the out of plane perturbations of one planet
affect the gap formation of the non--inclined planet? If both are on
inclined orbits, is the damping less efficient leading to longer
timescales? What happens when one planet is inside a disk cavity and
the other just outside of it?

\acknowledgments{We thank an anonymous referee for his useful comments
and suggestions. Parts of this work were carried out under the
auspices of the National Nuclear Security Administration of the U.S.
Department of Energy at Los Alamos National Laboratory under Contract
No. DE-AC52-06NA25396, for which this is publication number 
LA-UR~09-03432.}

{}

\end{document}